# Exam problems that link physical concepts with electrical phenomena in living cells.

Vadim Shlyonsky, Bertrand de Prelle, Florian Bodranghien, Don Patrick Bischop, and David Gall

Laboratoire d'Enseignement de la Physique, Faculté de Médecine, Université libre de Bruxelles, Bruxelles, Belgium

Abstract

We present a collection of biophysics examination problems that focus on electrodiffusion, the superposition principle, and currents in RC circuits, leading towards an understanding of the Hodgkin-Huxley model of action potentials. These problems link basic concepts in electricity to the physiology of living cells. The models described in these problems are intended help students develop an understanding of electrical processes in cells. Our work shows how introductory-level electricity courses can be taught in parallel with basic cell (electro)physiology and how basic concepts of electric circuits apply to cellular phenomena.

I. Introduction

In this paper we present a selection of examination problems that were used in a course on biophysics that was taught for more than a decade to undergraduate students at the Faculty of Medicine of the Université libre de Bruxelles. This course covers the physical/electrical principles underlying resting membrane phenomena and membrane excitation in living cells. It is taken by second-year premedical and biomedical sciences students who have already studied both introductory physics and calculus for one year, and a similar course could be appropriate for biomedical engineering students. The problem help illustrate the concepts that we present in this course. They may be of interest not only to instructors of similar courses, but also to instructors in more standard physics courses who wish to include problems with biological relevance.

It should be noted that there is no particular textbook that covers exactly all of the topics of the course; a specific textbook has been written for our students in French language, and is freely available upon request. The problems selected for this paper are arranged to follow the course syllabus. We begin with electrostatics and electrodiffusion and then advance to increasingly complex equivalent electrical circuits of biological membranes while progressively summarizing physiologically relevant electrical phenomena and concepts. The final aim of the course is to present the Hodgkin-Huxley model of cell excitability. This is a mathematical model in which cellular electrical phenomena are represented by a set of three nonlinear differential equations whose time-dependent solutions explain how electrical signals (i.e., action potentials) are generated in neurons.

The structure and complexity of these problems are configured to make the final examination a "learning by synthesis" according to Bloom's taxonomy.[1] During the semester, problems of similar complexity are assigned to students as open-book homework. Students are taught to draw sketches of problems or to complete already available ones to facilitate understanding. They can also meet instructors for additional guidance.

The overall goal of this type of learning is to show students that simple equivalent electrical circuits and basic knowledge of their functioning provides a foundation for modeling the natural processes governing resting and excitable cell membrane behavior. The term "excitable cell" describes a living cell whose plasma membrane contains voltage-dependent ion channels responsible for firing action potentials beyond a certain threshold membrane voltage value. As in any science, biophysical models are designed to match observations in the limits fixed by measuring instruments. To reinforce this, we offer learning-objective based comments on each problem in order to link its model with actually-measured examples of cellular electrical phenomena or with experimental approaches routinely employed in electrophysiology. To this end, many problems have been modified by introducing conceptual questions requiring answers in everyday language. These questions may be given as homework, but they are omitted on written examinations due to time limitations. The models for these exam questions are created to be as simple as possible (with the least number of components) while mimicking a broad range of natural phenomena. In other words, they are based on ideal conditions and their equivalent circuits, but do not contain details of real voltage/current clamp experiments such as those that lead to the artefacts related to Ag/AgCl-electrode potentials, liquid-junction potentials, or series resistance compensation, all of which are common issues in electrophysiology but not essential in our biophysics course. ("Clamp" here means that whole circuit is held either at a fixed voltage or fixed total current.) Consequently, these problems are not suitable for a pure electrophysiology class.

We hope that by becoming familiar with formulating cell behavior via simple electrical models, students will develop insights which will be useful for future applications of biophysics in medicine, biology, and biomedical technology. In our opinion, this is an original approach in life-sciences curricula as introductory physics and physiology are rarely taught together. This collection has been screened against available didactic literature on the same topic, and only original examples are included.[2-9] Less-relevant problems or problems of lower difficulty can be found in many excellent textbooks with end-of-chapter tutorial questions[10-23]. During examinations, students are given the following list of physical constants: $k = 1/(4\pi\varepsilon_0) = 9.0 \; 10^9 \; Nm^2C^{-2}$, $\mu_0 = 4\pi \; 10^{-7} \; TmA^{-1}$, $R = 8.314 \; Jmol^{-1}K^{-1}$, $F =$

$96500\ Cmol^{-1}$, $m_{proton} = 1.6726\ 10^{-27}$ kg, $q_{proton} = +1.601\ 10^{-19}$ C, and $N_A = 6.022\ 10^{23}$. These symbols will be used without definition in the problems below. Full analytic solutions of the problems can be found in the Supplementary Material.

## II. Problems

The selected problems are grouped into subsections, although these groupings are somewhat arbitrary due to the broad nature of the problems. Answers to some questions are given in [square brackets].

### II.1. Electrodiffusion

The following two problems demonstrate the practical application of differential expressions to find solutions of physics/physiology problems in simplified conditions. The learning goals are twofold. First, we want to have students understand that simple biophysical problems can be solved analytically and that graphical representations can help simplify the analyses. Second, we focus on negative signs in the differential equations. By convention, it is required to negate derivatives because slope values are always opposite to the direction of flux of positively charged particles. That is, the net flow of charged particles is in the direction of free energy minimization, i.e. from high to low concentration or voltage, so the mass flow is proportional to the negative derivative of the concentration/voltage. Since current is defined as the direction of positive charge flow, it is also proportional to the negative derivative of the concentration/voltage. These problems would be far more challenging if the equations are not given in the text.

*Problem 1.* In one dimension, the electrical current density $I_S$ (A/cm$^2$) corresponding to the movement of ions $S$ under the influence of a concentration gradient is given by Fick's law,

$$I_S = -z_S F D_S \frac{dC_S(x)}{dx}, \tag{1}$$

where $z_S$ is the valence of the ion $S$; $D_S$ is its diffusion coefficient; and $C_S(x)$ is the concentration of $S$ in one dimension. Consider a steady state where the number of Cl⁻ ions per liter of solution at $x = 0$ cm is $10^{23}$; this density decreases linearly to zero at $x = 10$ cm.

a. Calculate current density $I_S$ at $x = 2$ cm if the diffusion coefficient is $10^{-5}$ cm²/s. [-1.602·10⁻⁵ A/cm²]

b. What is the direction of electrical current under these conditions? In which direction do Cl⁻ ions move under these conditions?

c. Based on your answer to (b), what is the purpose of the negative sign in the Fick's law?

*Problem 2.* In one dimension, the electrical current density $I_S$ (A/cm²) corresponding to the movement of ions $S$ under the influence of electrical field is given by:

$$I_S = -\frac{F^2 z_S^2 D_S}{RT} C_S(x) \frac{dV(x)}{dx}, \tag{2}$$

where $z_S$ is the valence of the ion $S$, $D_S$ is its diffusion coefficient, $T$ is the absolute temperature, $C_S(x)$ is the concentration of $S$ at the point $x$, and $V(x)$ is the potential associated with the presence of an electrical field. Consider a stationary state where the only permeant ions present in the solution are calcium ions. The solution is homogenous, its concentration is 1 mol/L, and the temperature is 25°C. The electric potential at $x = 0$ cm is 10 V, and it decreases linearly to zero at $x = 5$ cm.

a. Calculate the current density $I_S$ at $x = 2$ cm if the diffusion coefficient is $10^{-5}$ cm²/s. [$0.3\ A/cm^2$]

b. What is the direction of electrical current under these conditions? In which direction do calcium ions move under these conditions?

c. Based on your answer to (b), what is the purpose of the negative sign in the differential equation?

Students next learn that transmembrane diffusion of ions can be modeled as diffusion in one dimension. That is electrodiffusion through biological membrane in the presence of both ionic gradient and electric field is described by the Nernst-Planck equation for one dimension, which is simply the combination of the diffusion equations in the above problems:

$$I_s = -D_s F z_s \left( \frac{dC_x(x)}{dx} + \frac{z_s C_s F \cdot dV(x)}{RT \cdot dx} \right) \tag{3}$$

Its analytic solution in the case of one permeant ion at electrochemical equilibrium (i.e. when the membrane current is zero) gives the Nernst equation and thus allows calculation of the electrochemical equilibrium potential of this ion:

$$E_s = \frac{RT}{z_s F} \ln \frac{C_{s,out}}{C_{s,in}} \tag{4}$$

In the case of several permeant ions, the solution to the Nernst-Planck equation leads, with some strong assumptions, to the resting (reversal) membrane potential equation, which is known to electrophysiologists as the Goldman-Hodgkin-Katz (GHK) equation [24]. Some of these assumptions are not entirely physically justified; for example, that the electric field within a biological membrane is constant, but are made for sake of simplicity. For solutions containing only monovalent cations M$^+$ and monovalent anions A$^-$ it takes the following form:

$$E_m = \frac{RT}{F} \ln \left( \frac{\sum_i^n P_{M_i^+}[M_i^+]_{out} + \sum_j^m P_{A_j^-}[A_j^-]_{in}}{\sum_i^n P_{M_i^+}[M_i^+]_{in} + \sum_j^m P_{A_j^-}[A_j^-]_{out}} \right) \tag{5}$$

A "permeability" factor ($P_{M_i^+}$ and $P_{A_j^-}$) is introduced, and students learn that the system, in which ion transport is governed by permeabilities expressed in cm/s, shows a rectifying behavior and thus deviates from linearity. However, under symmetrical ionic conditions (that is, where the concentration of each ion is the same on both sides of the membrane), the GHK current equation closely approximates Ohm's law and thus can eventually describe linear behavior of the membrane current.

*Problem 3.* Consider an electrically passive cell with ionic concentrations [Na$^+$]$_{ext}$ = 140 mM, [Na$^+$]$_{int}$ = 10 mM, [K$^+$]$_{int}$ = 150 mM, and [K$^+$]$_{ext}$ < 150 mM. The temperature is 25 °C.

 a. If the ion fluxes through the membrane of this cell in both directions (int-to-ext and ext-to-int) were calculated using permeability coefficients (Eq. 6 below), would the current through the membrane obey the Ohm's law under these conditions?

$$\Phi_s = P_s z_s^2 \frac{V_m F^2}{RT}[C_s] \qquad (6)$$

 b. If the cell membrane is permeable only to potassium ions and if the membrane potential is -75 mV, what is the extracellular potassium concentration? [$8.08\ mM$]

 c. If the membrane is permeable to both cations and the permeability ratio $P_K/P_{Na}$ is 100, find the extracellular concentration of potassium ions such that the observed membrane potential is depolarized by 10 mV compared to the membrane that is purely permeable to K$^+$ under the same conditions. [$\cong 2.94\ mM$]

Further ideas for electrodiffusion problems can be drawn from Benedek and Villars[2] (Chapter 3: problems 1-10, 23,26), Hobbie and Roth[7] (Section 6: problem 25 and Section 9: problems 14-17), and Parke[9] (Chapter 14: problems 3-5).

## II.2. Cellular RC circuits in stationary states.

 While electrodiffusion problems do not formally require an equivalent circuit diagram in order to be solved, such a diagram becomes essential in all following problems. The key model element of virtually any type of biological membranes is an RC-circuit as shown in Fig. 1.

This circuit contains membrane capacitance $C_m$ in parallel with different branches consisting of ionic conductances ($g_{Na}$, $g_K$, and leak conductance $g_l$) in series with their corresponding Nernst electrochemical equilibrium potentials ($V_{Na}$, $V_K$, and $V_l$). Depending on the problem, additional branches may be present, for example, to account for fluxes of the calcium or chloride ions. The membrane potential $V_m$ is determined as a difference between intracellular and extracellular potentials $V_{in}$ and $V_{out}$, with the latter set to zero by convention since the extracellular node is grounded. In some cases, students are forced to make a hypothesis about the direction of current flow in the circuit; this type of activity is one of our learning objectives. Students should be able to correctly interpret the text of the problem in order to identify (if it is not explicit) in which state the RC circuit is found. One possible state is a steady state where the membrane capacitor is fully charged and no further variation of membrane potential occurs. A second possible state is a pseudo-steady state occurring at the brief moment when the time derivative of the membrane voltage (*dV/dt*) changes sign, such as when the action potential peaks. Finally, we can have a non-steady state where the variations of potential on the membrane capacitor give non-zero capacitive currents, as these variations are superposed on the resting membrane voltage. The correct interpretation determines the correct direction of reasoning and hence the choice of formulae.

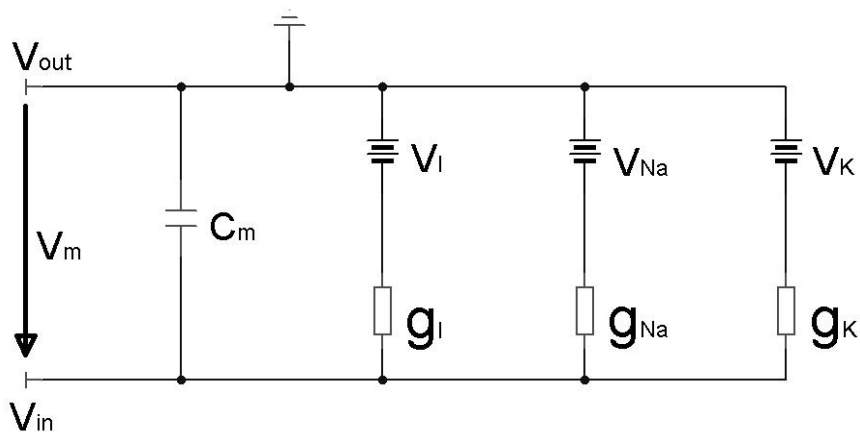

Figure 1. Equivalent RC-circuit of a biological membrane. It consists of membrane capacitance $C_m$ in parallel with the leak conductance $g_l$ and ionic conductances ($g_{Na}$, $g_K$).

Reversal potentials ($V_l$, $V_{Na}$, $V_K$) reflect ionic conditions of electrochemical equilibrium for each conductance branch. Membrane potential $V_m$ is defined as a difference between intracellular and extracellular electrical potentials ($V_{in}$-$V_{out}$). Since extracellular node is grounded $V_m$ equals basically to $V_{in}$.

In the following problem we begin by building a model of a biological membrane, which we represent with an equivalent circuit where the membrane capacitor is present but is not involved in solving the problem because the situation is steady-state. Here, students should demonstrate in literal form exactly how the resting membrane voltage (hence, in steady state) depends on the ionic gradients, which are represented by Nernst potentials *V₁* and *V₂*, and also on the corresponding membrane resistances governed by different ion channels. Students should understand that membrane potential is a result of a competition between conductances, where the highest conductance always wins and pulls the membrane voltage to the equilibrium value found in its circuit branch.

*Problem 4.* Figure 2 represents a cell membrane in a resting state; the membrane capacitor ($C_m$) is charged. The membrane contains two different ionic conductances (corresponding to resistances *R₁* and *R₂*) having different equilibrium potentials represented by voltage generators *V₁* and *V₂*.

    a. Calculate the stationary membrane capacitor potential *Vₘ* as a function of *V₁, R₁, V₂*, and *R₂*. $[V_m = \frac{V_2 R_1 + V_1 R_2}{R_1 + R_2}]$

    b. Show that if *R₁* is sufficiently greater than *R₂*, *Vₘ* tends to *V₂*, and that for *R₂* sufficiently greater than *R₁*, *Vₘ* tends to *V₁*.

    c. In what way does this problem resembles two voltage dividers?

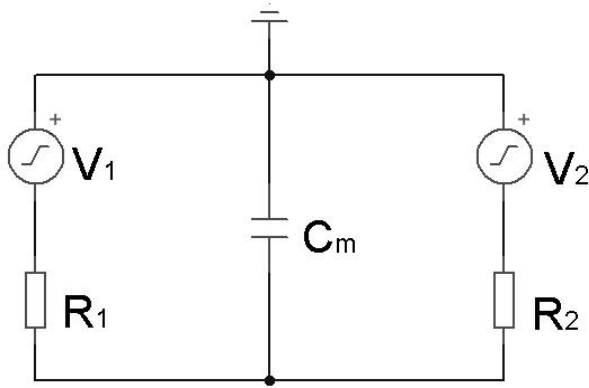

Figure 2. Circuit diagram of the problem 4.

We next continue with purely resistive circuits to demonstrate the laws of charge conservation and the principle of superposition. Here the learning objective is that the equivalent circuit of a single ion channel branch comprises a fixed resistor and a switch in series with a voltage generator that represents an ionic gradient, with all ion channels in the membrane connected in parallel. Students learn that by using a voltage clamp technique, one can fix any desired voltage across this circuit and, thus, quasi-instantly charge a membrane capacitor that is in parallel with a selected ion channel branch. The clamping condition here corresponds to fixing $V_{in} - V_{out}$ to a desired value. The electrophysiological recording appearing in problem 5 is taken from data acquired by students in one of our laboratory sessions[25-26].

*Problem 5.* Figure 3 shows an electrophysiological recording of gramicidin channel activity recorded at an applied voltage of -120 mV under symmetrical ionic conditions across a lipid bilayer membrane.

    a. Draw an equivalent electrical circuit for this bilayer. What is the maximal number of channels open at the same time in this bilayer?

b. Evaluate a unitary conductance of the channel observed and the leak resistance of the bilayer. [$R_L \cong 150$ GΩ, $\gamma$ is between 13.3 - 14.2 pS]

c. Make a series of graphs (one for each channel observed) that give the time evolution of unitary currents of each channel that would together correspond to the total ion channel currents observed. Align the time axis of your graphs with that of the given figure.

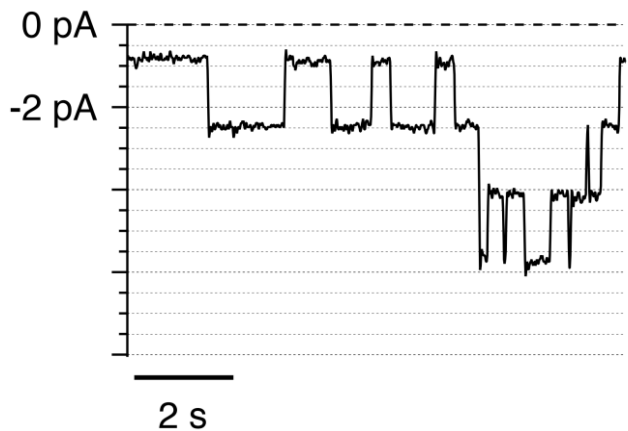

Figure 3. Gramicidin channels recording in a lipid bilayer membrane at -120 mV applied voltage and under symmetrical ionic conditions.

Now the capacitor becomes an active player in the circuit, increasing its complexity. The following problem tests knowledge of the corresponding electrostatics, namely the charge accumulated by a capacitor at a given voltage. Students already know that maintaining the cell membrane at a fixed potential induces a stationary state where Ohm's law describes the behavior of the voltage. Students should be able to make unit conversions and to deal with proportions; that is, we link geometry and physics.

*Problem 6.* Consider a spherical and electrically passive cell of membrane resistance 1.5 GΩ and specific membrane capacitance 1 µF/cm². The resting membrane voltage is -71 mV.

a. What current should be injected to maintain the cell at the potential of -65 mV? [$I_{inj}$ = 4 pA]

b. Under voltage-clamp conditions, we need to add an electrical charge of $30 \times 10^{-15}$ C to one side of the membrane to bring the potential to -80 mV. Calculate the radius of this cell. [$r = 5.15 \ \mu m$]

*Problem 7.* Consider a simplified neuron as shown in Fig. 4. The resting membrane resistance and capacitance of the soma are 1.2 GΩ and 30 pF, respectively. The cell is at a resting membrane voltage of -65 mV and the threshold for the appearance of action potentials in the soma is 10 mV above the resting potential. The dendritic membrane is electrically passive and has a space constant λ = 1 mm. Two stimulation electrodes are placed on two different dendrites at the same distance *d* from the soma.

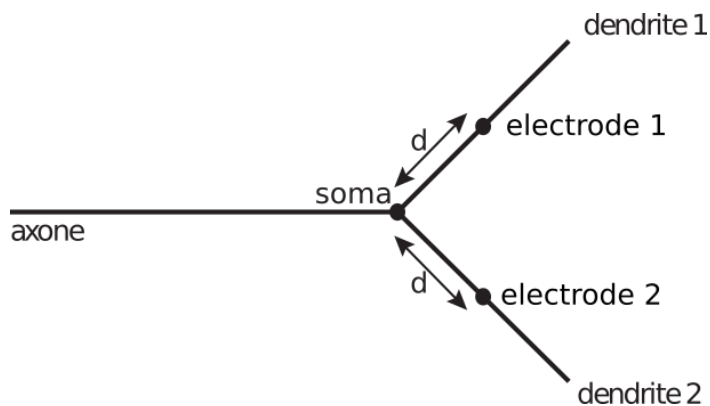

Figure 4. Schematical neuron. Stimulation electrodes are placed either on two dendrites at a distance *d* from soma or directly on soma.

a. If the current injected at each electrode locally depolarizes the membrane to -55 mV, at what maximum distance *d* can these electrodes be placed so that the membrane potential in the soma reaches the threshold of excitability during simultaneous injection of current by the two electrodes? [$d = 0.693 \ mm$]

b. In the absence of dendritic stimulation, calculate the duration of a 20 pA current pulse injected directly into the soma that brings the membrane potential to the threshold. [t = 19.4 $ms$]

Additional tutorial questions on biological applications of electrostatics and on RC circuits in stationary states can be found in Nelson et al.[3] (Chapter 12: problems 7, 8), Newman[4] (Chapter 15: problems 21, 22 and Chapter 16: problems 22, 23), Philips et al.[5] (Chapter 17: problems 3, 5), Hobbie and Roth[7] (Section 6: problems 27, 28, 32, 34-36, 52, 62, 66) and Herman[8] (Chapter 12: problems 11, 13, 15, 16, 19).

### II.3. Cellular RC circuits in non-stationary states

The following problem is based on Kirchhoff's current law and the superposition principle; that is, the resistive and the capacitive currents must be calculated separately and summed. Students should also recall that the sum of membrane currents is always zero in a resting cell.

*Problem 8.* Consider a spherical electrically passive cell whose membrane resistance and capacitance are 200 MΩ and 200 pF, respectively. The resting membrane voltage is -70 mV. This cell is voltage-clamped using a voltage protocol sketched in Fig. 5. Draw a graph giving the evolution of the total membrane current during this experiment.

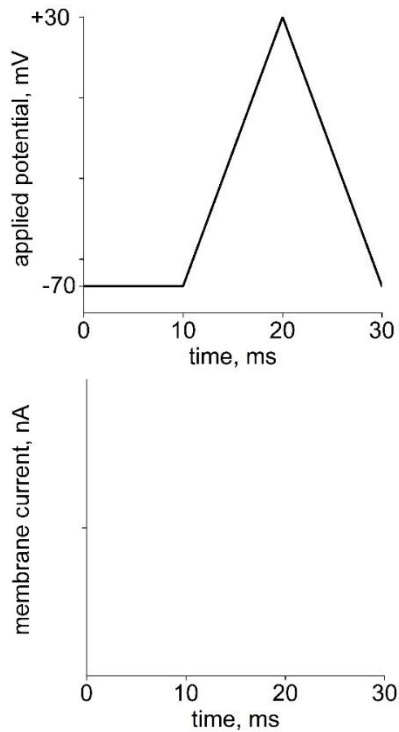

Figure 5. Voltage-clamp protocol applied in the problem 8 and an aligned graph for sketching corresponding evolution of the membrane current.

The learning objective of the following exercise is to show that the membrane potential of a cell that is permeable to only one type of ion does not depend on the membrane conductance. In other words, students study how a close-to-the-Nernst-type membrane approaches its steady state when the steady state depends only on the value of the Nernst potential in given conditions. This problem also demonstrates that the cell membrane behaves as a passive low-pass filter. This type of filter will have been seen during lectures.

*Problem 9.* Consider a spherical cell of a radius of 12 μm whose specific membrane capacitance is 1 μF/cm$^2$. The plasma membrane of this cell is permeable only to potassium ions, and these channels are not open all the time. This results in a specific conductance of 0.3 mS/cm$^2$. The intracellular and the extracellular concentrations of K$^+$ are 110 mM and 5 mM, respectively. The cell is at 37 °C. At t = 0 we instantly change the extracellular concentration of K⁺ to 50 mM.

    a. Qualitatively describe the evolution of the membrane potential starting at t = 0.

    b. Develop an equation that describes the evolution of the membrane potential starting at t = 0. $[V = V_{ini} + (V_{fin} - V_{ini}) \cdot (1 - e^{-300t})]$

    c. We now perform the same experiment in the presence of an agonist of potassium channels. This increases the time spent by the channels in the conductive state without any effect on single-channel conductances. How will the evolution of the membrane potential be modified compared to the situation in the absence of this agonist? [faster]

    d. What is the dependence of the membrane potential of this particular cell on the membrane conductance? [none]

    e. Based on your answers to the previous questions, can you explain in what way the cell membrane behaves like a low-pass filter?

    f. Based on your calculations in b, what do you think about the feasibility of such manipulation with extracellular medium exchange? [not feasible]

<u>Comment to the instructor:</u> Students should be able to solve the differential equation that starts with writing Kirchhoff's law of charge conservation for the intracellular node that is common between the capacitor and the resistor. The solution of this differential equation resembles an expression of the voltage relaxation in an RC-circuit.

As students advance, they learn about temporal and spatial signal summation, which will be used routinely in a later neurophysiology course. For the moment, these phenomena are studied using simple RC circuits. Knowledge of the superposition principle is required to solve the following two problems. Students separately treat different electrical phenomena, the charge and discharge of capacitor on top of a resting membrane voltage, and sum their contributions. The first problem demonstrates a practical approach in neural cell stimulation by current pulse trains in which a threshold potential value is attained by using pulse trains having short delays, in contrast to stimulation using a long single pulse. The second problem illustrates summation of synaptic currents at the space point corresponding to the soma of the neuron. In these problems we introduce thus for the first time the notion of a threshold potential beyond which the cell behavior is no longer passive.

*Problem 10.* Consider a spherical cell of radius of 11 μm. The specific membrane capacitance is 1 μF/cm². The membrane of this cell contains only ion channels perfectly selective for K⁺ ions, with a unitary conductance of 3 pS and a density of 0.5 channel per μm². These channels are permanently open. The cell resting membrane potential is -67 mV.

    a. At t = 0 we inject a first current pulse of constant amplitude 40 pA for 8 ms; see Fig. 6. What is the value of membrane potential observed at the end of this pulse? [-54.75 mV]

    b. After a time interval Δt during which no current is injected, a second current pulse of the same amplitude and duration is injected. What is the value of the membrane potential at the end of the second pulse if Δt = 1 ms and if Δt = 10 ms ? [-51.57 mV and -53.93 mV, respectively]

    c. Repeat the calculations for pulses of 40 ms duration. How do the results differ from the previous cases? [ΔV$_m$ = ΔV$_{max}$ in both cases]

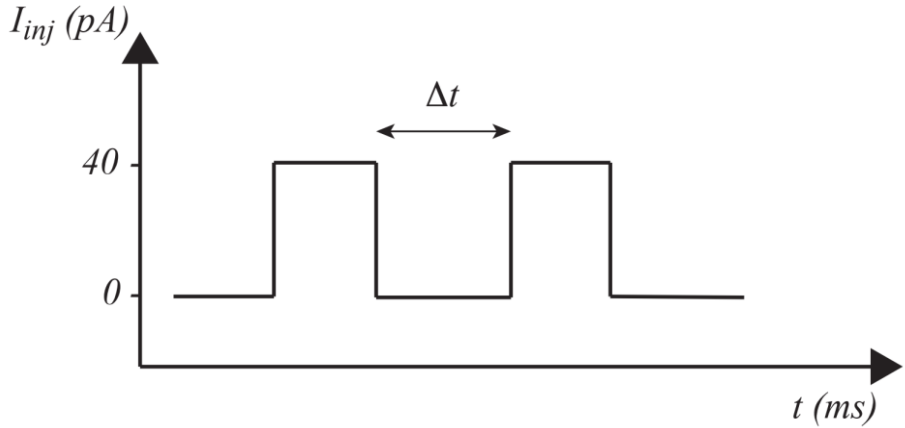

Figure 6: Current-clamp protocol used in the problem 10. Both current pulses are applied for either 8 ms or 40 ms. The time delay between two pulses is either Δt=1 ms or Δt=10 ms.

Comment to the instructor: In this problem we observe (1) the charging of the capacitor during the first pulse, (2) discharging of the capacitor during the time interval Δt and the time of the second pulse, and (3) the charging of the capacitor during the second pulse. All of these occur independently and on the top of the resting (i.e. stationary) membrane voltage generation. The time constant τ calculation requires accurate conversion of units and reasoning with proportions. When the value of τ is calculated, it becomes clear that there is no temporal summation when the pulses have duration of 40 ms because t > 5τ (recall that $e^{-5}$ < 0.01).

Several problems on RC-circuits in non-stationary states are presented in Hobbie and Roth[7] (Section 6: problems 29 and 37).

**II.4 Ionic currents, ion selectivity and excitability**

The learning objectives of the following two problems are to understand the important cellular biophysical parameters of the rheobase and chronaxie. The rheobase is the minimal amplitude of a rectangular current pulse of infinite duration that brings the membrane potential to the excitability threshold value, while the chronaxie is the minimal duration of a rectangular current pulse of the doubled rheobase amplitude after which the membrane potential reaches the excitability threshold value. Students need to appreciate that both of these depend only on the passive electrical properties of the cell: the leak resistance and membrane capacitance. Through this understanding, students realize that simple physics governs quite complicated physiological cell properties. The theoretical definitions of rheobase and chronaxie are required to solve these problems because they encode additional important information. When working these problems, we ask students to draw the equivalent circuit. Here again geometry links to physics.

*Problem 11.* Consider a spherical cell of diameter of 10 μm. At membrane potentials below the threshold of action potential generation (at $V_m < -54$ mV $= V_{th}$), this cell has ohmic behavior and its resistance R=1.5 GΩ is determined by the permanent opening of 10000 potassium-selective channels. The specific membrane capacitance is 1 μF/cm². The intracellular and the extracellular concentrations of K⁺ are 110 mM and 5 mM, respectively. The cell is at 37 °C.

a. What current should be injected to voltage-clamp this cell at -80 mV? In which direction do potassium ions move under these conditions? [$I_K = 1.73$ pA]

b. This cell is placed under osmotic conditions that linearly reduce its plasma membrane surface area by 50% in one hour; the cell remains spherical at all times. The ionic concentrations inside and outside the cell also remain the same. There is no exo- and endocytosis during this period; that is, the number of channels and their activities do not change. Draw a graph of the evolution of the rheobase and a graph of the evolution of the chronaxie during this hour. Include proper scales on your graphs. [$I_{rh} = 19.1$ pA = const; $\tau_{ch}$ decreases linearly from 3.26 ms to 1.63 ms]

*Problem 12.* Consider a spherical neuron whose membrane has a specific capacitance of 1 µF/cm². Its plasma membrane contains voltage-dependent potassium and sodium channels; the leak conductance is $g_L$ = 0.0125 mS/cm². The ionic conditions are such that the equilibrium potentials for potassium and sodium ions are -80 mV and 70 mV, respectively. The reversal potential of the leak conductance is -60 mV.

  a. Calculate specific sodium and potassium conductances ($g_{Na}$ and $g_K$) as well as their ratio in a resting cell at -67.2 mV if the total membrane resistance under these conditions is $R_{tot}$ = 3200 Ω·cm².[$g_{Na}$ = 0.025 mS/cm², $g_K$ = 0.275 mS/cm², $g_K/g_{Na}$ = 11]

  b. Calculate specific sodium and potassium conductances ($g_{Na}$ and $g_K$) as well as their ratio at the peak of the action potential of 41.7 mV, if the total membrane resistance under these conditions is $R_{tot}$ = 50 Ω·cm².[ $g_{Na}$ = 16.225 mS/cm², $g_K$ = 3.7625 mS/cm², $g_K/g_{Na}$ = 0.232]

  c. The permanent injection of -50 pA into the resting cell hyperpolarizes the plasma membrane by 12.74 mV. Calculate the radius of the neuron. [r = 10µm]

  d. Calculate the chronaxie of this neuron. [$\tau_{ch}$ = 2.218ms]

  e. What is the value of capacitive current in questions a, b and c? Explain. [$I_c$ = 0]

Physiologists can acquire much useful information about living cells from electrophysiological techniques like patch-clamp and lipid bilayer membrane measurements. Interpreting this information requires knowledge of several (bio)physical concepts, such as reversal potential and open probability of an ion channel, electrochemical driving force, the Nernst potential, and the Goldman-Hodgkin-Katz potential, all of which are discussed in the course. The term "ion selectivity" is also introduced.

*Problem 13.* Consider a spherical neuron with a radius of 15 μm; the plasma membrane has a specific capacitance of 1 μF/cm². This membrane contains 10000 ion channels with a unitary conductance of 2 pS. These channels are permeable to both $K^+$ and $Na^+$ with the selectivity ratio $g_K/g_{Na} = 4$. The ionic concentrations are $[Na^+]_{ext} = 140$ mM, $[Na^+]_{int} = 11$ mM, $[K^+]_{ext} = 5$ mM, and $[K^+]_{int} = 110$ mM. This experiment is performed at a temperature of 15 °C.

a. The application of an agonist increases the open probability of these channels from 0.4 to 0.8, without changing the selectivity ratio. Construct two I/V graphs showing the total membrane current as a function of the applied potential in the range between -100 mV and +100 mV in both the absence and presence of the agonist. Specify units on the axes. [$I = 20 \cdot P_O \cdot (V+48.7)$, where I is in pA and V is in mV]

b. Calculate the resting membrane potential of the cell in both the absence and presence of the agonist. [$V_{rev} = -48.7$ mV]

c. How does the open probability of the channels modulate the resting membrane potential of this cell?

*Problem 14.* A cell with negligible background membrane conductance expresses 2000 identical copies of ion channels in its plasma membrane. All we know is that these channels have unitary conductance of 15 pS, that they are constantly open, and are highly selective for one ionic species. During a voltage-clamp experiment performed at an applied potential of +30 mV, a total membrane current of 3.49 nA is measured. The ionic concentrations are $[Na^+]_{ext} = 140$ mM, $[Na^+]_{int} = 10$ mM, $[K^+]_{ext} = 7$ mM, $[K^+]_{int} = 100$ mM, $[Cl^-]_{ext} = 120$ mM, and $[Cl^-]_{int} = 4$ mM. The temperature is 22 °C.

   a. Determine for which ionic species these channels are selective. [Chloride]
   b. Which hypotheses did you have to make?

More ideas for this type of problems can be found in Hobbie and Roth[7] (Section 9: problems 19, 20-26).

**II.5. Hodgkin-Huxley model**

One of the ultimate goals of our course is to familiarize students with the Hodgkin-Huxley computational model of cell excitability. Readers can find a complete description of this model in Section 6.13 in Hobbie and Roth[7]. Exam questions on this model integrate practically all of the theory seen by students up to this point. For example, one of the learning objectives of the following problem is to show that the assumption of reaching 99% of the value of a parameter that follows exponential growth after exactly five time constants is not analytically correct. This problem describes an actual experiment performed by Hodgkin and Huxley when they were developing their model and needed to find the way to distinguish potassium and sodium ion currents.

*Problem 15*. We realize a voltage-clamp experiment using a squid giant axon at 8 °C in which the membrane potential is first clamped at -65 mV and then stepped to -20 mV.

   a. Given that the ionic concentrations are $[Na^+]_{int}$ = 50 mM, $[K^+]_{ext}$ = 20 mM, and $[K^+]_{int}$ = 400 mM, calculate the extracellular concentration of $Na^+$ for which the contribution of sodium channels to the membrane currents during this voltage step will be zero. [21.9 mM]

   b. In the Hodgkin-Huxley model of excitability, the kinetic functions $\alpha_n(V)$ and $\beta_n(V)$ that determine the opening of the voltage-dependent potassium channel are given by:

$$\alpha_n = \frac{-0.01 ms^{-1} mV^{-1}(55mV+V)}{e^{-0.1 mV^{-1}(55mV+V)} - 1} \tag{7}$$

$$\beta_n = 0.125 ms^{-1} e^{-0.0125 mV^{-1}(65mV+V)} \tag{8}$$

Calculate the fraction of potassium channels open in the stationary state at -20 mV.

[Po = 0.486]

c. What time will be needed to reach 99% of the stationary *n* value ($n_\infty$)? [t = 10.68 ms]

Other excellent tutorial questions about the Hodgkin-Huxley model are presented in Section 6 (problems 40-43, 48-50, 54) in Hobbie and Roth[7].

## II. Conclusions

The exam problems presented here require understanding of physical concepts and the application of both mathematics and physical laws. Their solutions are based on rigorous knowledge of the vocabulary of the physiology of living cells and knowledge of their excitable behavior; students need to be able to interpret special terminology in problem texts to extract pertinent data. We hope that the models described in these problems are of value to the students in that they will provide a foundation for them to develop their own insights. They may also be of value to teachers of biophysics as sources of new ideas for student instruction and assessment.

## III. References.

[1]B. S. Bloom, M. D. Engelhart, E. J. Furst, W. H. Hill and D. R. Krathwohl, *Taxonomy of educational objectives: The classification of educational goals. Handbook I: Cognitive domain*. (David McKay Company, New York, 1956).22

[2]G. B. Benedek and F. M. H. Villars, *Physics with Illustrative Examples from Medicine and Biology. Vol. 3. Electricity and Magnetism*, (Springer, New York, 2000).

[3]P. Nelson, M. Radosavljevic, S. Bromberg, *Biological Physics*, (W.H. Freeman & Co Ltd, New York, 2007).

[4]J. Newman, *Physics of the Life Sciences*, (Springer-Verlag, New York, 2008).

[5]R. Phillips, J. Kondev, J. Theriot, H. G. Garcia, *Physical Biology of the Cell*, 2$^{nd}$ ed. (Garland Science, London - New York, 2013).

[6]B. J. Roth and R. K. Hobbie, "A collection of homework problems about the application of electricity and magnetism to medicine and biology." Am. J. Phys. **82**, 422 (2014).

[7]R. K. Hobbie, B. J. Roth, *Intermediate Physics for Medicine and Biology*, 5$^{th}$ ed. (Springer-Verlag, New York, 2015).

[8]I. P. Herman, *Physics of the Human Body, Biological and Medical Physics, Biomedical Engineering*, (Springer International Publishing, 2016).

[9]W. C. Parke, *Biophysics: A Student's Guide to the Physics of the Life Sciences and Medicine*, (Springer International Publishing, 2020)

[10]J. B. Marion, *General Physics with Bioscience Essays*, 2$^{nd}$ ed. (John Wiley and Sons, New York, 1985).

[11]J. Kane and M. Sternheim, *General Physics*, (John Wiley and Sons, New York, 1991).

[12]P. Davidovits, *Physics in Biology and Medicine*, 3$^{rd}$ ed. (Academic Press, 2008).

[13]P. Nelson, *Biological Physics: Energy, Information, Life*, 1$^{st}$ ed. (W. H. Freeman and Company, New York, 2008).

[14]S. A. Kane, *Introduction to Physics in Modern Medicine*, 2$^{nd}$ ed. (CRC Press, 2009).

[15]K. Franklin, P. Muir, T. Scott, L. Wilcocks and P. Yates, *Introduction to Biological Physics for the Health and Life Sciences*, (John Wiley and Sons, New York, 2010).

[16]K. A. Dill and S. Bromberg, *Molecular Driving Forces: Statistical Thermodynamics in Biology, Chemistry, Physics, and Nanoscience*, 2nd ed. (Garland Science, London - New York, 2011).

[17]I. D. Campbell, *Biophysical Techniques*, (Oxford University Press, 2012).

[18]A. W. Wood, Physiology, Biophysics, and Biomedical Engineering, (Series in Medical Physics and Biomedical Engineering, CRC Press, 2012).

[19]P. R. Kesten and D. L. Tauck, *University Physics for the Physical and Life Sciences: Volume II*, 1st ed. (W. H. Freeman, 2012).

[20]T.A. Waigh, *The Physics of Living Processes: A Mesoscopic Approach*, (John Wiley and Sons, 2014).

[21]R. D. Knight, B. Jones, S. Field, *College Physics: A Strategic Approach*, 3rd ed. (Pearson Education Limited, 2015).

[22]F. Cleri, *The Physics of Living Systems*, (Springer International Publishing, Switzerland, 2016).

[23]J. S. Walker, *Physics*, 5th ed. (Pearson Addison-Wesley, San Francisco, 2017)

[24]B. Hille, *Ion channels of excitable membranes*, 3rd ed. (Sinauer Associates, Sunderland, 2001)

[25]V. Shlyonsky, F. Dupuis, D. Gall, "The OpenPicoAmp: An Open-Source Planar Lipid Bilayer Amplifier for Hands-On Learning of Neuroscience." PLoS ONE, **9**, e108097 (2014).

[26]V. Shlyonsky, D. Gall, "The OpenPicoAmp-100k: an open-source high-performance amplifier for single channel recording in planar lipid bilayers." Pflugers Arch, **471**, 1467-1480 (2019).